\documentclass[]{aastex631}
\usepackage{amsmath}
\usepackage{CJK}
\usepackage{booktabs}
\usepackage{threeparttable}
\usepackage{amsthm,amsmath,amssymb}
\usepackage{mathrsfs}
\usepackage{gensymb}

\shorttitle{xing et al}
\shortauthors{xing et al}
\graphicspath{{./}{figures/}}

\begin{document}
\begin{CJK*}{UTF8}{gbsn}

\title{Simulating the Escaping Atmosphere of GJ436 b with two-fluid Magnetohydrodynamic Models \footnote{Revised Manuscript on Dec, 15, 2023}}

\email{guojh@ynao.ac.cn}

\author[0000-0001-7655-0920]{lei xing (邢磊)}
\affiliation{Yunnan Observatories, Chinese Academy of Sciences \\
P.0.Box110 \\
Kunming, 650216, People’s Republic of China}
\affiliation{University of Chinese Academy of Sciences \\
No.19(A) Yuquan Road, Shijingshan District \\
Beijing, 100049, People’s Republic of China}
\affiliation{Key Laboratory for Structure and Evolution of Celestial Objects, Chinese Academy of Sciences \\
P.0.Box110 \\
Kunming, 650216, People’s Republic of China}
\affiliation{International Centre of Supernovae, Yunnan Key Laboratory\\
            Kunming 650216, P. R. China\\}

\author[0000-0002-8869-6510]{jianheng guo (郭建恒)}
\affiliation{Yunnan Observatories, Chinese Academy of Sciences \\
P.0.Box110 \\
Kunming, 650216, People’s Republic of China}
\affiliation{University of Chinese Academy of Sciences \\
No.19(A) Yuquan Road, Shijingshan District \\
Beijing, 100049, People’s Republic of China}
\affiliation{Key Laboratory for Structure and Evolution of Celestial Objects, Chinese Academy of Sciences \\
P.0.Box110 \\
Kunming, 650216, People’s Republic of China}
\affiliation{International Centre of Supernovae, Yunnan Key Laboratory\\
            Kunming 650216, P. R. China\\}

\author[0000-0001-6903-5306]{chuyuan yang (杨初源)}
\affiliation{Yunnan Observatories, Chinese Academy of Sciences \\
P.0.Box110 \\
Kunming, 650216, People’s Republic of China}

\author[0000-0002-8519-0514]{dongdong yan (闫冬冬)}
\affiliation{Yunnan Observatories, Chinese Academy of Sciences \\
P.0.Box110 \\
Kunming, 650216, People’s Republic of China}
\affiliation{University of Chinese Academy of Sciences \\
No.19(A) Yuquan Road, Shijingshan District \\
Beijing, 100049, People’s Republic of China}
\affiliation{Key Laboratory for Structure and Evolution of Celestial Objects, Chinese Academy of Sciences \\
P.0.Box110 \\
Kunming, 650216, People’s Republic of China}
\affiliation{International Centre of Supernovae, Yunnan Key Laboratory\\
            Kunming 650216, P. R. China\\}


%
%
%



\begin{abstract}
Observations of transmission spectra reveals that hot Jupiters and Neptunes are likely to possess escaping atmospheres driven by stellar radiation. Numerous models predict that magnetic fields may exert significant influences on the atmospheres of hot planets. Generally, the escaping atmospheres are not entirely ionized, and magnetic fields only directly affect the escape of ionized components within them. Considering the chemical reactions between ionized components and neutral atoms, as well as collision processes, magnetic fields indirectly impact the escape of neutral atoms, thereby influencing the detection signals of planetary atmospheres in transmission spectra.In order to simulate this process, we developed a magneto-hydrodynamic multi-fluid model based on MHD code PLUTO. As an initial exploration, we investigated the impact of magnetic fields on the decoupling of H$^+$ and H in the escaping atmosphere of the hot Neptune GJ436 b. Due to the strong resonant interactions between H and H$^+$. The coupling between them is tight even if the magnetic field is strong. However, our simulation results indicate that under the influence of magnetic fields, there are noticeable regional differences in the decoupling of H$^+$ and H. With the increase in magnetic field strength, the degree of decoupling also increases. For heavier particles such as O, the decoupling between O and H$^+$ is more pronounced. our findings provide important insights for future studies on the decoupling processes of heavy atoms in the escaping atmospheres of hot Jupiters and hot Neptunes under the influence of magnetic fields.
\end{abstract}



\section{Introduction} \label{sec:intro}
As of 2023, more than 5500 exoplanets have been discovered (data sourced from exoplanet.eu), comprising approximately 8\% classified as hot Jupiters and hot Neptunes, with orbital distances from their host stars of less than 0.1 AU. Exploring exoplanets has opened up new possibilities for the discovery of extraterrestrial life. Currently, much of the data on exoplanets is obtained from space telescope projects such as HST, CHEOPS, TESS, JWST, and others. In the near future, Chinese missions like CHES\citep{Ji2022} and Earth 2.0\citep{Ge2022}, based on space telescopes, will provide a new outlook in the search for  Earth twins. Furthermore, research on planetary atmospheres plays a crucial role in exploring key questions such as planetary evolution, the origin of life, and planetary habitability\citep{Zhang2020}. \cite{Vidal2003} provided the first confirmation of the expanded atmosphere in the hot Jupiter HD209458 b through observations of Ly$\alpha$ transmission spectra. Additionally, it was established that hydrogen atoms in the atmosphere could reach escape velocities of up to 100 km/s\citep{Vidal2003, Holmstrom2008}, a speed comparable to that of the solar wind. The one-dimensional atmospheric model proposed by \cite{Yelle2004, Murray2009, Guo2016} reveals that this signal likely originates from a hydrodynamic escaping atmosphere heated by stellar radiation. The signal of H escaping at 100 km/s is strongly indicative of charge exchange processes occurring between the planetary and the stellar wind \citep{Holmstrom2008, Khodachenko2017}. Subsequent simulations of the atmospheres of multiple exoplanets, such as HD189733 b \citep{Rumenskikh2022}, WASP-52 b\citep{Yan2022}, GJ436 b\citep{Khodachenko2019}, and GJ3470 b\citep{Shaikhislamov2021}, have further confirmed the existence of such hydrodynamic escape atmosphere on these planets. This observation suggests that hydrodynamic escaping atmospheres are prevalent on hot Jupiters and hot Neptunes. Moreover, the occurrence of intense hydrodynamic escape may be closely associated with the formation of short-period hot Neptune deserts\citep{Owen2017}. 

Similarly, the hot Neptune GJ436b exhibits a strong Ly$\alpha$ absorption signal, with blue wing absorption reaching up to 70\%, and a noticeable asymmetry in the absorption profile between the red and blue wings \citep{Ehrenreich2015, Lavie2017, Kulow2014}. \cite{Khodachenko2019} elucidates that such observational results are intimately linked to the generation of energetic neutral atoms (ENAs) through the charge exchange between stellar and planetary wind. Subsequent observations in other wavelengths of the hot Jupiter HD209458 b have also revealed the presence of elements such as C, O, Mg and Si in the escaping atmosphere \citep{Vidal2004, Vidal2013, Ballester2015, Linsky2010}. However, in the case of GJ436b, current observations of other elements such as C, O, N, and Si exhibit relatively weak signals \citep{dosSantos2019}. Recently, \cite{Rumenskikh2023} proposed that the absence of helium observation signals may be attributed to the limited distribution range of metastable helium and the impact of radiation pressure on metastable helium. The causes of the absence of transmission spectrum signals for other elements remain to be investigated. Considering the observed absorption signals of oxygen (O) on several exoplanets, such as HD209458 b \citep{BenJaffel2007, BenJaffel2008} and HD189733 b \citep{BenJaffel2013}, the decoupling of H$^+$ and O is also worthy of discussion.

Additionally, magnetic fields represent a significant factor influencing atmospheric escape on planets. According to \cite{BenJaffel2022}, a comprehensive analyses were conducted by combining multi-band observations of the HAT-P-11 planetary-star system with holistic simulations of the atmosphere from the lower atmosphere (approximately 200 bar) up to the magnetosphere. As a result, it was estimated that the magnetic field strength in the equatorial region of this exoplanet should be in the range of 1-5 Gauss. \cite{Trammell2014} elucidated that under the influence of a magnetic field, planetary atmospheres can give rise to distinct features such as wind-zones and dead-zones. Furthermore, \cite{Khodachenko2021} demonstrated that a magnetic field as low as 1 Gauss could restrict atmospheric escape, forming a dead-zone particularly in regions proximate to the equator of the planet. Under the influence of stellar XUV radiation, the atmosphere undergoes ionization, and the thermalization process of photoelectrons heats the atmosphere, triggering hydrodynamic escape. In the absence of considering magnetic fields, \cite{Hunten1987, Zahnle1986, Guo2019} modeled the possibility of decoupling phenomena between H and other heavy atoms in the escaping atmospheres of Earth-like planets. Additionally, \cite{Xing2023} demonstrated the potential decoupling of helium and other particles in the atmosphere of the hot Jupiter HD209458 b. Typically, the escaping atmospheres of hot planets are partially ionized plasmas, where the charged constituents directly interact with the magnetic field, while neutral atoms do not. This can lead to the decoupling of charged particles and neutral particles under the influence of the magnetic field. Simultaneously, it may impact particles that have already undergone decoupling under the influence of gravity.

To investigate this decoupling process, we developed a magneto-hydrodynamic multi-fluid program based on the well-established  MHD code PLUTO \citep{Mignone2007, Mignone2012}. As a starting point, we explored the potential decoupling of H and H$^+$ in the escaping atmosphere of the hot Neptune GJ436 b, considering the influence of magnetic fields on the decoupling process. The model incorporates two fluids: one representing H and the other comprising ions resulting from the combination of H$^+$ and electrons. The inclusion of ion-electron coupling is motivated by the strong Coulomb forces between ions and electrons. The paper will be structured into the following segments: Section \ref{sec:model_description} presents our magneto-hydrodynamic multi-fluid model; Section \ref{sec:results} outlines the computational results of the model; Section \ref{sec:Discussions} engages in a discussion of the findings and Section \ref{sec:Conclusions} provides a summary of the study.

\section{Multi-fluid MHD model} \label{sec:model_description}
To investigate the potential decoupling effects between H and H$^+$ in the expanded escaping atmosphere of the hot Neptune GJ 436b and the influence of a global magnetic field on such decoupling, we constructed a two-dimensional multi-fluid magnetohydrodynamic model. Referring to the planetary and stellar configurations described in \cite{Khodachenko2015, Trammell2014}, we averaged the atmosphere along the azimuthal direction. For simplicity, we neglected the planet's rotation, resulting in zero components of velocity and magnetic field in the $\phi$ direction. Consequently, both the velocity and magnetic field are treated as vector fields in the radial and polar directions. The computational code used in the model development is originated from our 2023 code \citep{Xing2023}. Our construction of the multi-fluid magnetohydrodynamic (MHD) model involves dismantling PLUTO first and then reconstructing its HD and MHD modules to adapt them for multi-fluid model computations. The subsequent Appendix \ref{sec:apdx} validates that the reconstructed MHD calculation module is functioning properly. This approach allows different components in the atmosphere to be simulated using either the hydrodynamic module or the magneto-hydrodynamic module based on their charge quantity. The interactions between components were calculated using the semi-implicit method described by \cite{Garcia2007}. This enables our model to simulate the escape processes of multi-component atmospheres separately and investigate decoupling phenomena among them.

\subsection{The multi-fluid HD equations}\label{sec:HD_equations}
We have employed a dual-fluid model for hydrogen atoms and hydrogen ions, where the strong Coulomb interaction between electrons and ions results in the combination of electrons and ions into a charged ionic fluid, still referred to as the H$^+$ fluid. The specific equations solved are as follows:

\begin{equation}\label{eq:ctn_1}
\frac{\partial n_ {t}}{\partial t}+\nabla\cdot(n_ {t}\textbf{u}_ {t}) = 0,\ t=n,s
\end{equation}

\begin{equation}\label{eq:n_mmt_2}
\frac{\partial (m_{n}n_{n}\textbf{u}_{n})}{\partial t} + \nabla\cdot(m_{n}n_{n}\textbf{u}_{n}\textbf{u}_{n} + \textbf{I}p_{n})
 = C_{ns}m_{amu}n_{n}n_{s}(\textbf{u}_{s}-\textbf{u}_{n})+m_{n}n_{n}\textbf{a}_{ext}
\end{equation}

\begin{equation}\label{eq:s_mmt_3}
\frac{\partial (m_{s}n_{s}\textbf{u}_{s})}{\partial t} + \nabla\cdot(m_{s}n_{s}\textbf{u}_{s}\textbf{u}_{s} 
 + \textbf{I}(p_{s}+\frac{B^2}{8\pi}) -\frac{1}{4\pi}\textbf{BB}) 
= C_{ns}m_{amu}n_{n}n_{s}(\textbf{u}_{n}-\textbf{u}_{s})+m_{s}n_{s}\textbf{a}_{ext}
\end{equation}

\begin{equation}\label{eq:prs_4}
\frac{\partial p_{t}}{\partial t}+
(\textbf{u}_{t}\cdot\nabla)p_{t}+\gamma p_{t}(\nabla\cdot\textbf{u}_{t})                                                                                                                                                                                                                                                                                                                                                                                       = 0,\ t=n,s
\end{equation}

\begin{equation}\label{eq:magf_5}
\frac{\partial \textbf{B}}{\partial t}=\nabla\times(\textbf{u}_{s}\times\textbf{B})
\end{equation}

Here, $n$, $\textbf{u}$, $p$, and $\textbf{B}$ represent the number density, velocity, pressure, and magnetic induction in (c.g.s) units, respectively. $m_{amu}$ denotes the atomic mass unit, and $C_{ns}$ is the collision parameter for H and H$^+$. Although we use the term "collision" here, the actual physical process involves resonant charge exchange between H and H$^+$, resulting in momentum and energy exchange, see details in \cite{Xing2023}. Subscripts `n' and `s' represent H and H$^+$, respectively. Equation \ref{eq:ctn_1} represents the continuity equation for H and H$^+$. In our model, we haven't included any source terms. Instead, we have set a fixed value for H$^+$/H at the bottom boundary, and the initial conditions also reflect this value. Equations \ref{eq:n_mmt_2} and \ref{eq:s_mmt_3} represent the momentum equations for H and H$^+$. Equation \ref{eq:prs_4} represents the evolution equation for the pressure of H and H$^+$. Here a polytropic index of $\gamma = 1.0001$ is used to simulate an isothermal atmosphere. Equation \ref{eq:magf_5} represents the evolution equation for the magnetic field. The external force $\textbf{a}_{ext}$ is adopted following the approach in \cite{Trammell2014}.

\begin{equation}\label{eq:aext}
\textbf{a}_{ext} = -\nabla[-\frac{GM_p}{r}-\frac{G(M_\star+M_p)r^2}{2D^3}(fsin^2\theta-1)]
\end{equation}

The planetary and stellar parameters involved in the model, i.e., the planetary mass $M_p$, planetary radius $R_p$, stellar mass $M_\star$, and the semi-major axis of the planetary orbit $D$, are obtained from the `exoplanet.eu' database. Referring to the discussion in \cite{Trammell2014}, due to the current neglect of planetary rotation in our model, the factor $f$ here should be taken as 3/2. Compared to Xing et al. (2023), our model's major breakthrough lies in the addition of a magnetohydrodynamic (MHD) module, enabling the computation of magnetic field evolution and its impact on the mulit-fluid model. As a starting point for MHD model calculations, we currently omit the ionization heating effects of stellar XUV radiation, Ly$\alpha$ cooling, and complex chemical reactions presented in the Xing et al. (2023) fluid model. Instead, we adopt an isothermal gas model to represent the heating effects. Additionally, we do not differentiate between electron and ion temperatures and do not separately solve for the electron temperature.

\subsection{Model settings}\label{sec:model_setting}
On the computational grid, we employed a spherically symmetric two-dimensional grid with the z-axis as the magnetic axis, assuming that the z-axis is perpendicular to the planetary orbital plane. The $\phi$ direction was averaged, and considerations were made for the grid in the $r$ and $\theta$ directions. In the $r$ direction, an exponentially stretched grid from the innermost layer with a width of $10^{-3}R_p$ to the outermost layer with a width of 10$R_p$ was adopted. The range of $r$ is from 1$R_p$ to 10$R_p$, with 400 grid points and a ratio of 1.01 between the adjacent grids. In the $\theta$ direction, a uniform grid with 256 points was used.

In our model, the equations of the continuity, momentum and pressure of H were solved by utilizing the hydrodynamic computation module from PLUTO. On the other hand, the three equations for H$^+$ and the evolution of the magnetic field \textbf{B} were computed using the magnetohydrodynamic module in PLUTO. The solvers used for both flows were HLL which estimate the propagation speed of the fastest wave at the discontinuity, leading to a two-wave model for the structure of the exact solution.\citep{Toro1994}. The Courant number (CFL) was set to 0.4, considering that the two flows have their respective time steps, and the minimum value between the two was chosen. 
For controlling $\nabla \cdot \textbf{B}$, we employed the eight-wave method provided by PLUTO \citep{Mignone2012, Mignone2007}.

The model includes four boundaries: $r=R_p$, $r=10R_p$, $\theta=0$, and $\theta=\pi$. At $r=R_p$, the density ($\rho$) and pressure ($p$) for H and H$^+$ were fixed, and $\textbf{u}=0$. Setting the lower boundary $\textbf{u}=0$ only affects a very small portion of the atmosphere (a few grid points), which promotes numerical stability. It has almost no impact on the final atmospheric structure, except for the few grid points near the lower boundary, while the rest of the domain satisfies flow conservation. The heat source driving atmospheric escape is the heating of the isothermal atmosphere, which counteracts the cooling effect of adiabatic expansion. the impact of $\textbf{u}=0$ on atmospheric escape can be neglected. At $r=10R_p$, outflow boundary conditions were set, while at $\theta=0$ and $\theta=\pi$, reflective boundary conditions were applied.

\section{Results} \label{sec:results}
\begin{table}[!t]
\begin{center}
\begin{threeparttable}[t]
\centering
\caption{Parameters for the models}

\begin{tabular}{lllllll}
\toprule
  Model &T(K) &$\rho_0(10^{-14}g/cm^3)$ &$B_0$(Guass) &$I_0(ion\ rate)$ &$\beta_0$ &$\dot{M}(10^{10}g/s)$\\

\midrule
  1 &3000 &$5$ &1 &0.5 &1.866 &2.95\\
  2 &3000 &$5$ &2 &0.5 &0.467 &2.62\\
  3 &3000 &$5$ &3 &0.5 &0.207 &2.22\\
  4 &4000 &$5$ &2 &0.5 &0.622 &15.8\\
  5 &3500 &$5$ &2 &0.2 &0.435 &1.90\\
  6 &3500 &$5$ &2 &0.5 &0.544 &7.90\\
  7 &3500 &$5$ &2 &0.8 &0.653 &16.3\\

\bottomrule

\end{tabular}\label{tab:para_list}
From left to right, the three columns represent the model number, lower boundary temperature T, lower boundary density $\rho_0$, magnetic pole field strength $B_0$, ionization rate $I_0$, plasma $\beta_0$ at the magnetic pole, and mass loss rate.

\end{threeparttable}
\end{center}
\end{table}

\begin{figure}[htbp]
\plotone{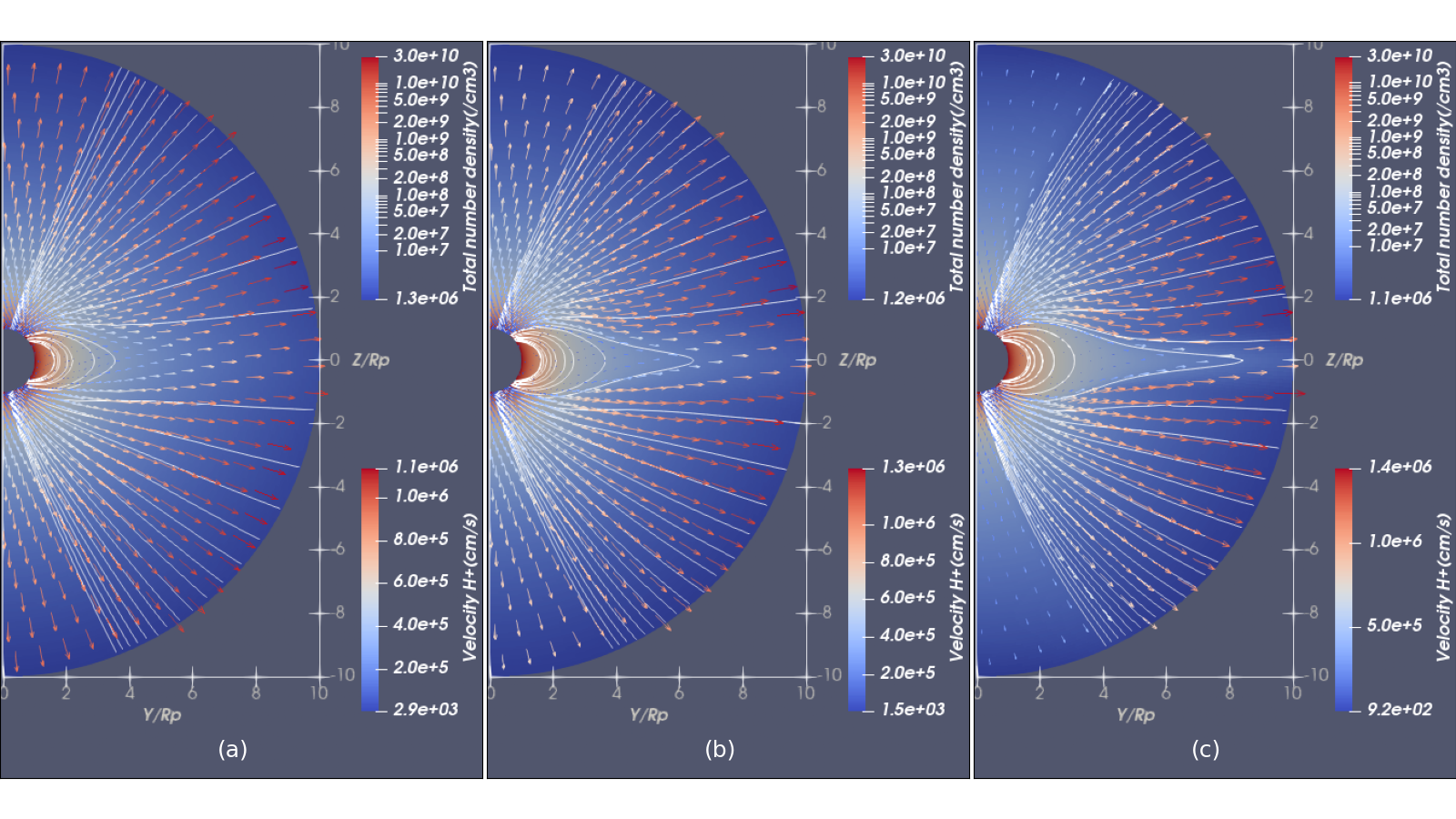}
\caption{These three subplots illustrate the changes in atmospheric structure with varying magnetic fields (models 1, 2, 3 from Table \ref{tab:para_list}). At a temperature of 3000K and an ionization rate of 0.5, as the magnetic field strength $B_0$ increases from 1G to 3G, the total number density of H and H$^+$ is depicted in the color map. Given the tightly coupled velocity vectors of H and H$^+$, only the vector field of H$^+$ is presented here. The vector size is represented by arrow length and color, with white lines indicating magnetic field lines.
}
\label{fig:b1_5}
\end{figure}

\begin{figure}[htbp]
\plotone{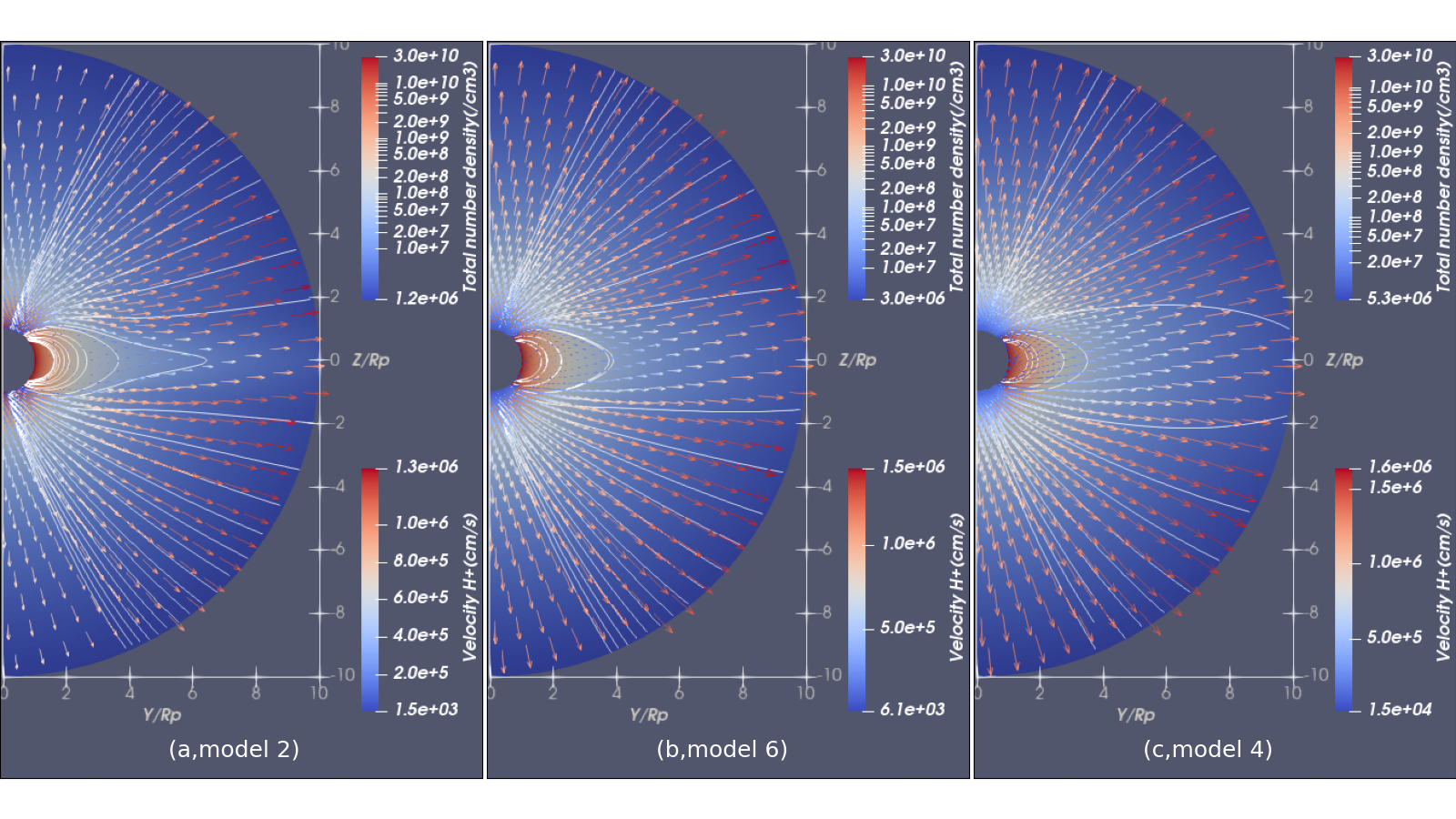}
\caption{
These three subplots depict the variations in atmospheric structure with changes in temperature (models 2, 6, 4 from Table \ref{tab:para_list}). At $B_0=2G$ and an ionization rate of 0.5, as $T_0$ increases from 3000 K to 4000 K, the total density of H and H$^+$ is illustrated in the color map. The vector field represents the velocity of H$^+$, with arrow length and color indicating vector magnitude. White lines denote magnetic field lines.
。}
\label{fig:T3_5k}
\end{figure}

\begin{figure}[htbp]
\plotone{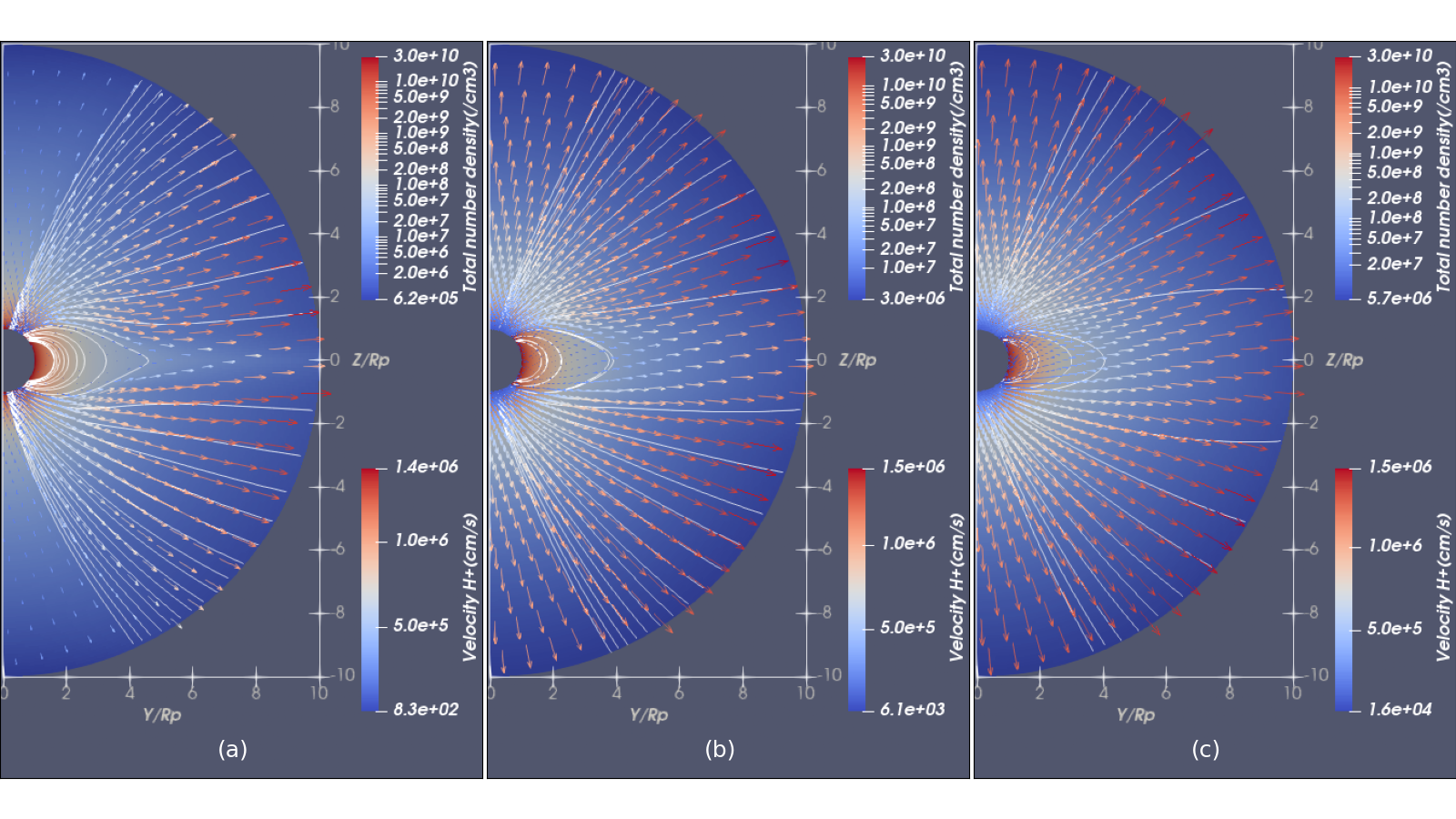}
\caption{These three subplots illustrate the variations in atmospheric structure with changes in ionization rate (models 5, 6, 7 from Table \ref{tab:para_list}). At $T_0=3500$ K and $B_0=2$ G, the total density of H and H$^+$ is depicted in the color map. The vector field represents the velocity of H$^+$, with arrow length and color indicating vector magnitude. White lines denote magnetic field lines.}
\label{fig:I20_90}
\end{figure}

The planetary parameters for GJ436 b are sourced from the exoplanet.eu website. In our model, the following parameters have been set: temperature $T_0$ (ranging from 3000 K to 4000 K), the density at the lower boundary $\rho_0$ ($5 \times 10^{-14} \text{g/cm}^3$), the magnetic field at the magnetic pole $B_0$ (ranging from 1 G to 3 G), and the ionization rate of H at the lower boundary $I_0$ (ranging from 0.2 to 0.8). The temperature range is based on references such as \cite{Salz2016}, indicating that the atmospheric temperature of this hot Neptune varies around 3000-4000 K. At these temperatures and densities, the pressure ranges from 0.02 to 0.04 $\mu$ bar, slightly above the homopause position.

The parameters for several computational results are displayed in Table \ref{tab:para_list}. The plasma $\beta_0$ at the lower boundary polar region is calculated using the formula:
\begin{equation}
\beta_0 = \frac{8\pi p_0}{(B_0/2)^2}
\end{equation}

The parameter $p_0$ represents the pressure at the lower boundary of the atmosphere. Due to the high coupling between the number densities of H and H$^+$ at the lower boundary, the magnetic field's influence on H$^+$ motion will also affect H motion (as will be discussed later in our discussion on particle decoupling). Conversely, the motion of neutral gas will also affect the evolution of the magnetic field. Therefore, $p_0$ here represents the sum of the pressures of H and H$^+$. Figure \ref{fig:b1_5} illustrates the impact of the magnetic field on the atmospheric structure. From left to right, $B_0$ is 1, 2, and 3 G, respectively. In these examples, a Dead Zone in the equatorial region is evident. With the increase of the magnetic field, there is a noticeable enhancement in the control effect of the magnetic field on the fluid, as observed through the comparison of the fluid velocity field and the magnetic field. These results are consistent with \cite{Trammell2014}. Additionally, as the magnetic field increases, the fluid velocity in the polar region shows a clear decreasing trend, eventually leading to the appearance of a dead zone in the polar region. Comparing the substance loss rates given in the table, it can be seen that as the magnetic field increases, the substance loss rate decreases, indicating a certain inhibitory effect of the magnetic field on fluid escape. However, there is an evident antagonistic trend between gas pressure and magnetic pressure, especially near the Dead Zone in the equatorial region, as seen in Figure \ref{fig:T3_5k}. With increasing temperature, the gas pressure gradually counteracts the magnetic pressure, and the Dead Zone controlled by the magnetic field tends to shrink. The substance loss rate also increases substantially, as shown by Table \ref{tab:para_list}. The fluid in the polar region, which was approaching disappearance, gradually recovers. If the density, temperature, and magnetic field are kept constant while increasing the ionization rate, the increased ionization rate will cause an increase in the number density of the fluid. With constant temperature, the gas pressure will also increase, leading to an increase in $\beta$ and a weakening of the magnetic field's control over the fluid. The results are similar to the case of a direct increase in temperature (Figure \ref{fig:I20_90}).

\begin{figure}[htbp]
\plotone{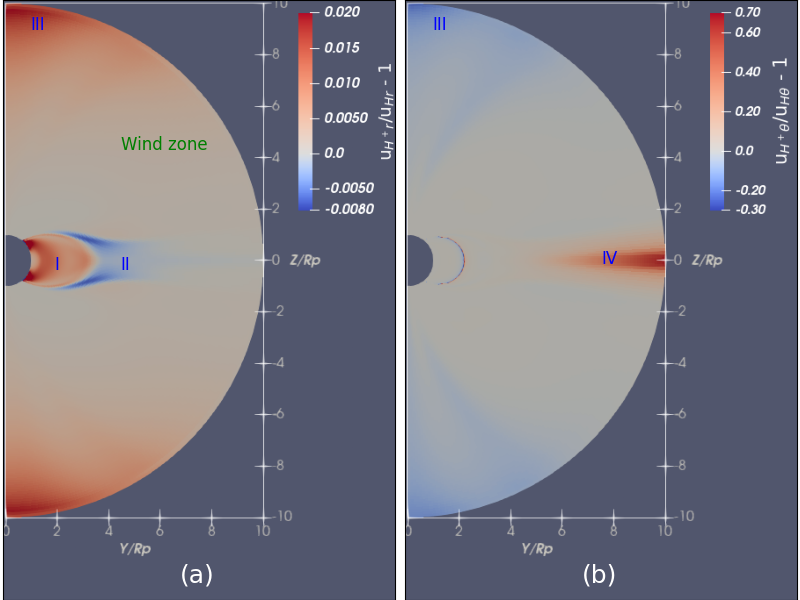}
\caption{These two subplots illustrate the decoupling of H and H$^+$ in the case of Model 5 from Table \ref{tab:para_list}. In plot (a), the radial velocity difference between H and H$^+$ is depicted, with the color map representing $u_{H^+r}/u_{Hr}-1$. In plot (b), the difference in $\theta$-direction velocities between H and H$^+$ is shown, with the color map representing $u_{H^+\theta}/u_{H\theta}-1$. "I" to "IV" and "Wind zone" denote different regions of the escaping atmosphere.}
\label{fig:case5_vdiff}
\end{figure}

\begin{figure}[htbp]
\plotone{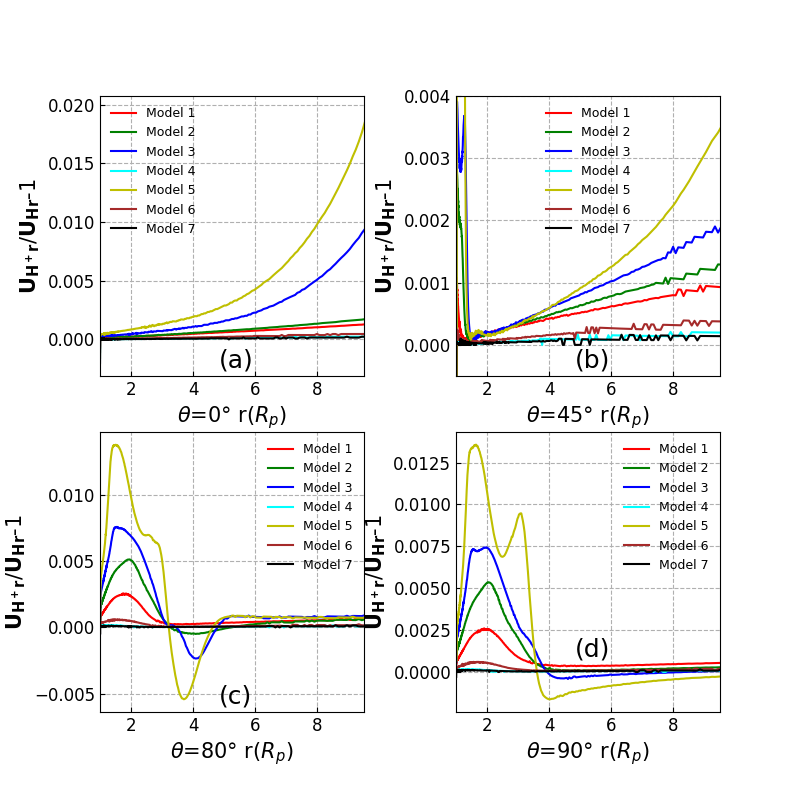}
\caption{
This figure compares the decoupling situation of the radial velocity components of H$^+$ and H at different $\theta$ angles (0$\degree$, 45$\degree$, 80$\degree$, 90$\degree$) from Table \ref{tab:para_list}. The four subplots present the difference in velocities between H$^+$ and H, i.e., $u_{H^+r}/u_{Hr}-1$.
}
\label{fig:vrdiff}
\end{figure}

Next, we are more concerned about the decoupling of H and H$^+$. As shown in Figure \ref{fig:case5_vdiff}a, the radial velocity decoupling in model 5 is not very apparent, with the difference in velocities between H and H$^+$ being less than 2\%. The degree of decoupling can be roughly divided into four regions. The degree of decoupling in the wind zone is relatively low because it is the main fluid escape area, and the magnetic field direction here is basically consistent with the fluid direction (refer to Figure \ref{fig:I20_90}a), resulting in a relatively small hindrance of the magnetic field to fluid escape. The degree of decoupling between H$^+$ and H is approximately 0.4\% in this region, refer to Figure \ref{fig:vrdiff}(b).

In region III, where the binding effect of the magnetic field on the fluid is small, it is noteworthy that under isothermal conditions, since the H$^+$ flow contains both H$^+$ and e$^-$, the pressure per unit mass of H$^+$ will be twice as much as of H. Therefore, H$^+$ will escape more easily than H, creating a dragging effect on H. The degree of decoupling in the $\theta$ direction seems to be slightly larger than in the radial direction. In the region where the magnetic field gradually loses control, labeled as II in Figure \ref{fig:case5_vdiff}, due to the direct limitation of H$^+$ escape by the magnetic field without a direct impact on H, hydrogen atoms will have a dragging effect on hydrogen ions. As a result, the velocity of hydrogen atoms will be slightly greater than that of hydrogen ions.

\begin{figure}[htbp]
\plotone{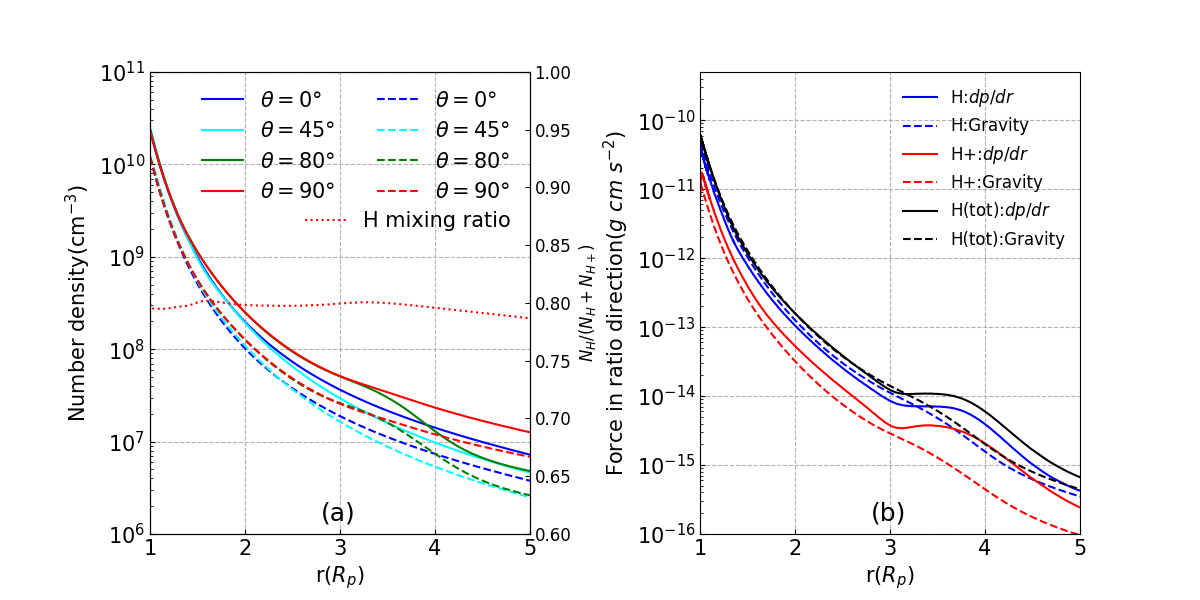}
\caption{
These two panels (a) and (b) respectively depict the number density structure and force distribution of H and H$^+$ for model 5 in Table\ref{tab:para_list}. In panel (a), solid lines represent the number density structure of H at four polar angles (0\degree, 45\degree, 80\degree, 90\degree), while corresponding dashed lines depict the number density structure of H$^+$. The red dotted line represents the distribution of the mixing ratio of H in polar angle $\theta = 90\degree$. In panel (b), the solid and dashed blue lines represent the pressure gradient force and gravitational force acting on H, respectively. Similarly, the red lines represent the pressure gradient force and gravitational force acting on H$^+$ and black lines represent the total pressure gradient force and total gravitational force.
}
\label{fig:case5_f_aly}
\end{figure}

In equatorial dead zone, it was expected that due to the constraint of the magnetic field, $u_{H^+r}$ should be smaller than $u_{Hr}$, but the numerical simulation results showed the opposite. Consequently, an analysis of the density structures and force situations of H and H$^+$ in this region was conducted. As shown in Figure \ref{fig:case5_f_aly}(a), the density structures of H and H$^+$ at various polar angles indicated that in the polar region and the dead zone near the equator (0\degree and 90\degree), the density was slightly higher than in the wind zone (45\degree). Additionally, due to the strong coupling between these two particles, their mixing rates remained almost constant. The dotted line in the figure represents the mixing rate of hydrogen when the polar angle is 90\degree. Referring to the analysis of the equatorial dead zone in \cite{Trammell2011}, equatorial dead zone in a quasi-fluid static equilibrium state, where the total pressure gradient force and total gravity of the fluid are balanced, as shown by the solid and dashed black lines in Figure \ref{fig:case5_f_aly}(b), and the total gravity force and total pressure gradient force are almost balanced. However, under isothermal conditions, the pressure of hydrogen ions of equal mass is twice that of hydrogen atoms, but the density structures of H and H$^+$ are almost proportional(dotted red line in Figure\ref{fig:case5_f_aly}(a)). This implies that the pressure gradient force of H is redundant and keeps H in quasi-fluid static equilibrium through drag. The force analysis of H$^+$ and H in panle (b) precisely confirms this inference. In the dead zone region, it can be observed that the gravitational force of H is slightly greater than the pressure gradient force, while for H$^+$, it is the opposite. Through the drag effect of H$^+$ on H, the difference between the gravitational force of H and the pressure gradient force of H is precisely compensated. Naturally, to form a drag effect, the velocity of H$^+$ needs to be greater than that of H, which can also be observed in similar phenomena in the subsequent test case of decoupling H$^+$ and O.

\begin{figure}[htbp]
\plotone{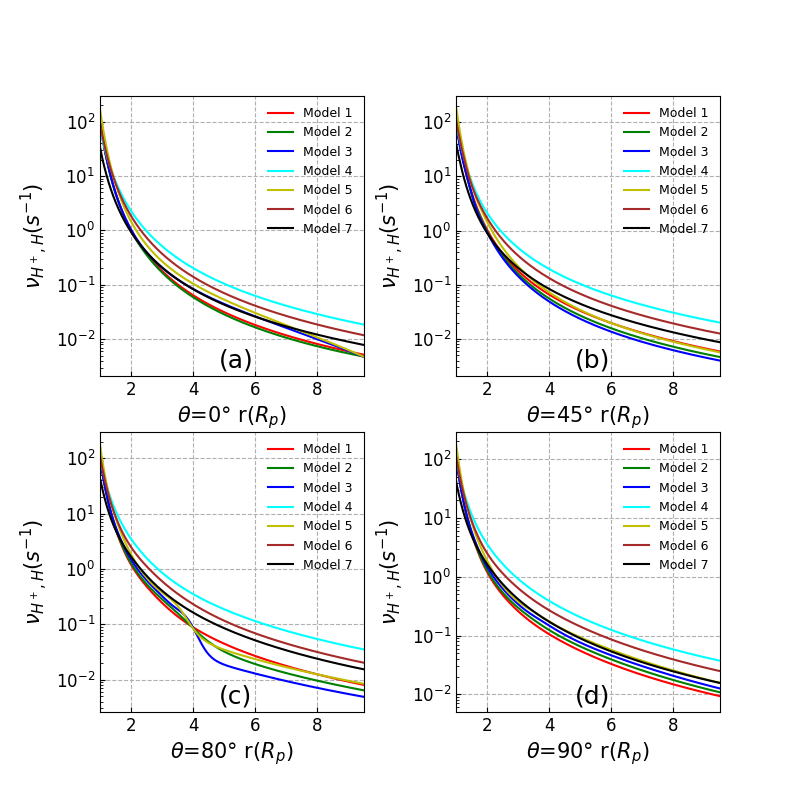}
\caption{
This figure shows the ion-neutral resonance interaction (charge exchange) frequency of H with H$^+$ at four polar angles (0$\degree$, 45$\degree$, 80$\degree$, 90$\degree$), as listed in Table \ref{tab:para_list}. $\nu_{H^+,H}$ is given by $\nu_{H^+,H} = 2.65\times 10^{-10}n_{H}T_r^{1/2}(1-0.083log_{10}T_r)^2$ according to \cite{Schunk1980}, where $T_r$ is average temperature of H$^+$ and H.
}
\label{fig:col_aly}
\end{figure}

In model 5, it can be noted that there are two regions with strong decoupling, namely the polar region and the equatorial region, with the decoupling in the polar region ($\approx 2\%$) even stronger than that in the equatorial region ($\approx 1.3\%$) . The rough decoupling region in the polar region is between $7-10 R_p$, while in the equatorial region, it is roughly between $1-3 R_p$. The particle number density in the polar decoupling region is orders of magnitude lower than that in the equatorial dead zone, resulting in a much lower $\nu_{H^+, H}$ as depicted in Figure \ref{fig:col_aly} (a) and (b).  Therefore, decoupling is easier in the polar region where collision frequencies are lower. The degree of decoupling in the $\theta$ direction seems to be slightly larger than in the radial direction. Decoupling is mainly observed in regions III and IV in Figure \ref{fig:case5_vdiff}b, while the decoupling in the equatorial dead zone and wind zone in the $\theta$ direction appears negligible.

Subsequently, let’s compare the decoupling situations under different parameters. 
In Figure \ref{fig:case5_vdiff}, the decoupling conditions in different regions of the escaping atmosphere exhibit significant differences. As a result, we will discuss three regions: the polar region, the wind region, and the equator. As shown in Figure \ref{fig:vrdiff}(a), models 3 and 5 exhibit noticeable decoupling characteristics in the polar region, with model 5 reaching a degree of decoupling of up to 2\% for H and H$^+$. From Figure \ref{fig:b1_5}(c) and Figure \ref{fig:I20_90}(a), it can be observed that models 3 and 5 have formed relatively distinct dead zones in the polar region compared to other models. We tentatively conclude that the decoupling in the polar region is closely related to the formation of a dead zone. In these two models with significant decoupling, the velocity of H$^+$ in this region is slightly greater than that of H. This is because the pressure exerted by H$^+$ flow per unit mass is twice that of H, making H$^+$ more prone to escape and exerting a dragging effect on H.

In the wind zone, as shown in Figure \ref{fig:vrdiff}(b) at $\theta = 45\degree$, the degree of decoupling increases linearly with radius in all 7 models roughly. However, the degree of decoupling here is relatively small compared to the dead zone in the polar region, reaching up to 0.3\%. The slightly faster escape of H$^+$ compared to H is for the same reason as in the polar region. H$^+$, containing H and e$^-$, has higher pressure and thus is more prone to escape. By comparing with the column of mass loss rates in Table \ref{tab:para_list}, we find that the degree of decoupling is inversely related to the mass loss rates, indicating an anti-correlation between the two quantities. We suspect that higher material loss rates will increase the coupling degree between H and H$^+$. This would result in maintaining a stable mixing ratio of H and H$^+$ in the escaping atmosphere, as shown by the red dotted line in Figure\ref{fig:vrdiff}(a). By comparing Figures \ref{fig:b1_5}, \ref{fig:T3_5k}, and \ref{fig:I20_90}, it can be observed that the final escape velocity does not vary significantly, around 15 km/s. According to mass conservation, an increase in atmospheric material loss rate corresponds to an increase in number density. In the wind region (Figure \ref{fig:col_aly}(b)), the structures of collision frequency for H$^+$ and H $\nu_{H^+,H}$ in these models are quite similar and can be correlated with the mass loss rates listed in Table \ref{tab:para_list} too. Additionally, here we provide $\nu_{H^+,H}$ for polar angles of 0\degree, 80\degree, and 90\degree for reference, as shown in Figure \ref{fig:col_aly}(a), (c), and (d). Similarly, the collision frequency and mass loss rate in these regions exhibit a certain degree of correlation. In models 4, 6, and 7 with higher mass loss rates, the collision frequency $\nu_{H^+,H}$ between H and H$^+$ is generally higher. However, the sequence of models 4, 6, and 7 may not correspond directly to the mass loss rates due to changes in the mixing ratio of H in the atmosphere, which also affects the collision frequency $\nu_{H^+,H}$.

For the equatorial region, the decoupling situation is more complex compared to the previous two regions. We consider two polar angles (80\degree, 90\degree), where 90\degree represents the decoupling situation at the equator, and 80 \degree focuses on the variation of decoupling from the dead zone to the wind zone in the atmosphere. Figure \ref{fig:vrdiff}(d) presents the decoupling situation at the polar angle of 90\degree. Comparing with Figures \ref{fig:b1_5}, \ref{fig:T3_5k}, and \ref{fig:I20_90}, it is evident that models 5, 3, 2, and 1 with pronounced equatorial dead zones exhibit stronger decoupling, indicating that equatorial decoupling depends on the formation of equatorial dead zones. Among these models, 1, 2, and 3 control other variables unchanged, with only the magnetic field varying. The increase in magnetic field corresponds to a gradual increase in decoupling strength. Comparing models 2 and 5 with the same magnetic field, model 5 exhibits less material loss, resulting in stronger decoupling. This suggests that decoupling in dead zones may depend on the relative relationship between atmospheric material loss and magnetic field. A stronger magnetic field is more likely to cause fluid decoupling, but the increase in atmospheric mass loss rate plays a suppressing role in decoupling.

\begin{figure}[htbp]
\plotone{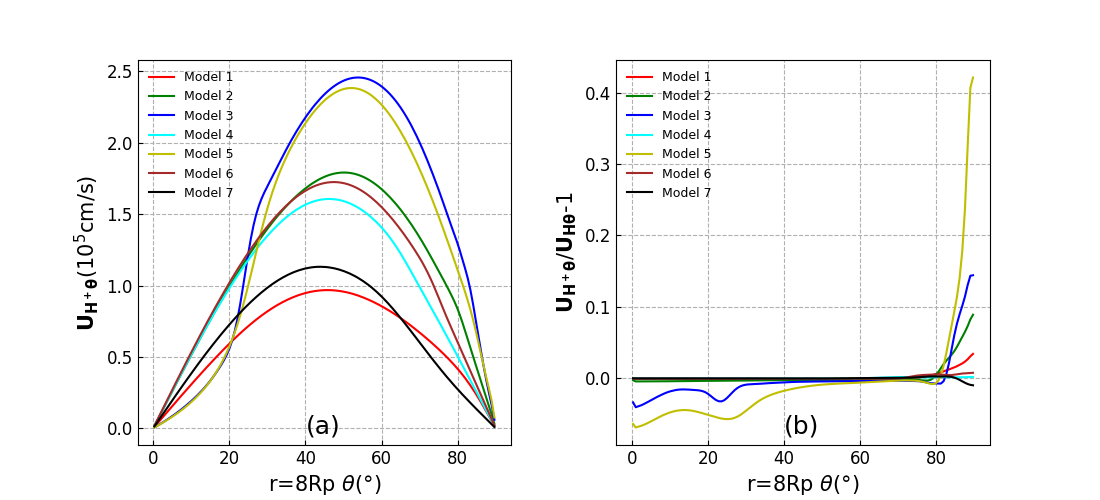}
\caption{This figure presents a comparison of the decoupling situation of the $\theta$-component of particle velocities from Table 1. The left plot illustrates the $\theta$-component of H$^+$ velocity at 8$R_p$, while the right plot shows the difference in $\theta$-component velocities between H$^+$ and H at 8$R_p$, i.e., $u_{H^+\theta }/u_{H\theta}-1$.
}
\label{fig:vtdiff}
\end{figure}

Additionally, we find a decoupling reversal phenomenon in the equatorial region. In models 3 and 5, as shown in Figure \ref{fig:vrdiff}(c), in the region of 1-3Rp, a quasi-fluid hydrostatic equilibrium zone forms within the dead zone. To maintain a similar mixing ratio to H, H$^+$ needs to counterbalance some of the gravitational force acting on the H flow through drag. Consequently, the velocity of H$^+$ is slightly higher than that of H. In the region of approximately 3-5Rp, as the fluid gradually moves away from fluid control, hydrogen atoms become more prone to escape due to the magnetic field only restricting ion escape. However, this reversal does not always occur. For model 1 with weaker magnetic fields, such reversal did not happen, resulting only in a slight weakening of the degree of decoupling. Conversely, as depicted in Figure \ref{fig:vtdiff}(a), the velocity component in the $\theta$ direction is two orders of magnitude smaller than the central radial velocity component. In some regions, the degree of decoupling in the $\theta$ direction may exceed that of the radial velocity component, reaching the order of 40\%. Decoupling mainly occurs in the polar and equatorial regions far from the planet. Due to the decrease in particle number density, decoupling tends to increase gradually (see Figure \ref{fig:case5_vdiff}b).

\section{Discussions}\label{sec:Discussions}

\begin{figure}[htbp]
\plotone{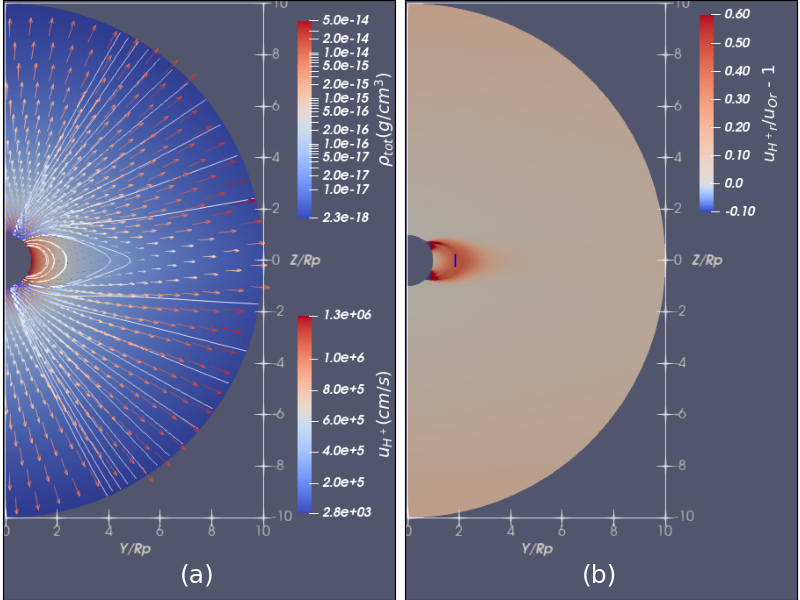}
\caption{Testing the case of proton(H$^+$) and oxygen (O) collisions, the model remains largely unchanged. Hydrogen atoms are replaced by oxygen atoms, with a temperature of 4300 K. The mass ratio of oxygen (O) and ionized hydrogen (H$^+$) at the lower boundary is set to 1:1. The magnetic field at the magnetic pole is 1 G. Subplot (a) displays the atmospheric structure, with the total density represented by a color map. The vector field represents the velocity of H$^+$, with arrow length and color indicating vector magnitude. White lines denote magnetic field lines. Subplot (b) illustrates the difference in radial direction velocities between these two particles, with the color map representing $u_{H^+r}/u_{Or}-1$. I denotes the equatorial dead zone in the model, where a pronounced decoupling between O and H$^+$ is observed.
}
\label{fig:HII_O}
\end{figure}

\begin{figure}[htbp]
\plotone{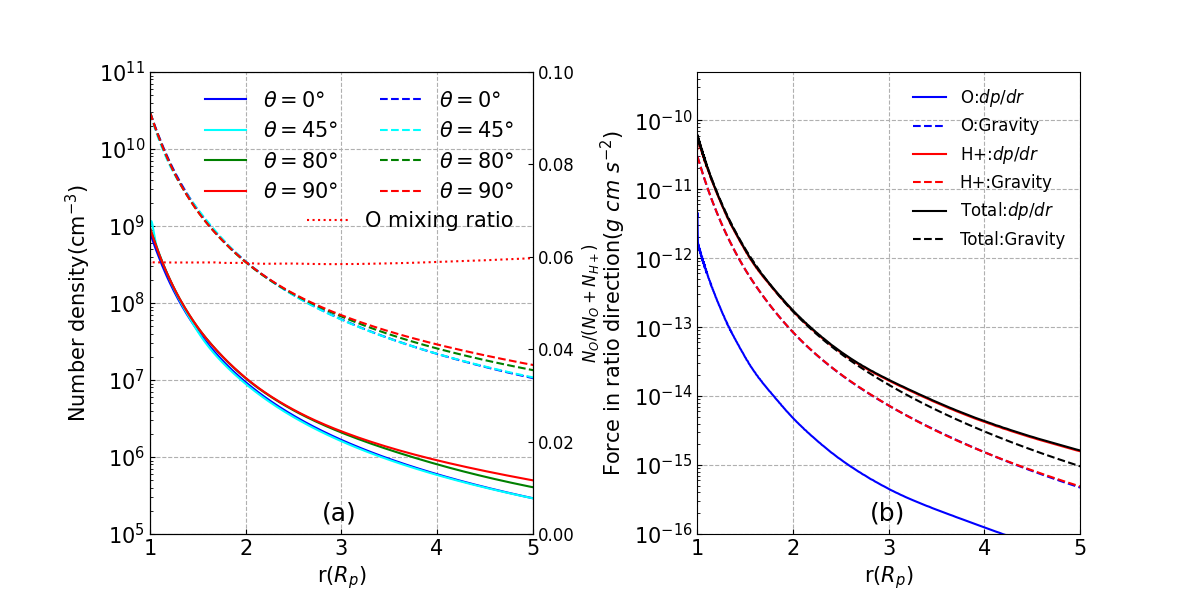}
\caption{Similar with Figure\ref{fig:case5_f_aly}, these two panels (a) and (b) respectively depict the number density structure and force distribution of H$^+$ and O for test model described in Figure\ref{fig:HII_O}. In panel (a), solid lines represent the number density structure of O at four polar angles (0\degree, 45\degree, 80\degree, 90\degree), while corresponding dashed lines depict the number density structure of H$^+$. The red dotted line represents the distribution of the mixing ratio of O in polar angle $\theta = 90\degree$. In panel (b), the solid and dashed blue lines represent the pressure gradient force and gravitational force acting on O, respectively. Similarly, the red and black lines represent the total pressure gradient force and gravitational force acting on H$^+$ and both species combined, respectively. In this model, the mass ratio of O to H$^+$ is 1:1 at lower boundary. Consequently, the gravitational forces acting on O and H$^+$ almost overlap (red and blue dashed lines). In such a scenario, the particle number density of the O flow is much smaller than that of H$^+$, resulting in the pressure of the O flow being significantly lower than that of H$^+$. As a result, the total thermal pressure gradient force and the thermal pressure gradient force of H$^+$ almost coincide (black and red solid lines).}
\label{fig:case_Of_aly}
\end{figure}

Our current model indicates a relatively small degree of decoupling between H$^+$ and H, with some cases reaching only a fraction of a percent, such as in models 4, 6, and 7. The limited decoupling is largely attributed to the high resonant charge exchange frequency between H and H$^+$ \citep{Schunk1980}. Similar phenomena are observed in the multi-fluid simulations of hydrogen-rich atmospheres in \cite{Guo2011} and \cite{Xing2023}, where the coupling between H and H$^+$ is tight, and even the models with the maximum decoupling in \cite{Guo2011} show only around 10\% with mass loss rate less than 10$^9g/s$ in hot Jupiter HD209458 b.

However, for non-resonant charge exchange processes, such as those between H and He$^+$, or considering binary diffusion between neutral particles, the frequency could be much lower. Multi-fluid simulations targeting these particles might exhibit more pronounced decoupling, as seen in \cite{Xing2023}, where there is a notable decoupling phenomenon observed for helium atoms from other particles. Additionally, \cite{Guo2019} has revealed the decoupling processes of O, H, and ions in the atmosphere. The inclusion of magnetic field interactions further complicates the decoupling scenario. As a supplement, let us discuss the decoupling situation of O and H$^+$ here. We replaced hydrogen atoms in the model with oxygen atoms and applied collision frequencies of H$^+$ and O \citep{Schunk1980}. An example is provided here, and as shown in Figure \ref{fig:HII_O}, the decoupling of H$^+$ and O is significantly stronger compared to the less pronounced decoupling between H$^+$ and H. The radial velocity decoupling can reach up to 50\% in dead zone. As shown in Figure \ref{fig:case_Of_aly}b, the gravitational forces within the dead zone in the equatorial region are primarily balanced by the pressure gradient force of H$^+$. To maintain the unchanged mixing ratio of O caused by high collision rates (Figure \ref{fig:case_Of_aly}a), H needs to provide additional drag force. Similar to the analysis for Model 5, this situation results in the velocity of H$^+$ being greater than that of O. Of course, the extreme ratio of $N_{H^+}/N_{O}$ at 16:1, as opposed to the solar ratio of $N_{H}/N_{O}$ at 1:0.0009\citep{Asplund2009}, is considered. This provides a reference for future studies on the decoupling of heavy particles in the atmosphere.

Given the diverse signals observed in the observations of hot Jupiters and hot Neptunes, including elements like He, C, O, Si, etc., a complex multi-fluid model is likely necessary to simulate the signals of these particles more accurately. Moreover, such a multi-fluid model is crucial for investigating the indirect interactions between magnetic fields and these particles. Our test cases further illustrate the significance of adopting this approach.

On the other hand, our current magnetohydrodynamic multi-fluid model is relatively simple, adopting an isothermal model with a fixed atmospheric ionization level. In reality, the escape of the atmosphere is driven by stellar radiation, particularly the XUV (X-ray and extreme-ultraviolet) radiation from the star, which ionizes components in the atmosphere, generating photoelectrons. These photoelectrons then contribute to atmospheric heating and escape. This process involves complex photochemical reactions, leading to temperature and ionization structures that can be different from our current model.

Moreover, the stellar wind is another crucial factor. Stellar wind not only affects the outer structure of the planetary atmosphere but also has a substantial impact on observational signals, such as Lyman-$\alpha$ observations. The charge exchange between stellar wind and planetary wind contributes significantly to the generation of energetic neutral atoms (ENAs) \citep{Holmstrom2008, Khodachenko2017}.

In our magnetohydrodynamic model, we currently neglect the effect of magnetic diffusion, implying that the electrical resistivity in the fluid is considered to be zero. However, in reality, the electrical resistivity of the ionized atmosphere is a factor that influences the magnetic field structure.

In our current model, we have not taken into account the rotation of the planet. However, for hot Jupiters and hot Neptunes, which are typically in close proximity to their host stars, tidal locking is a common assumption. Considering rotation, especially in the context of tidal locking, could give rise to global-scale circulation patterns \citep{Trammell2014}. To investigate this process, a minimum requirement would be a two-dimensional model with three vector components for velocity and magnetic field. However, since our model involves two vector components, addressing this complexity is one of the directions for future model improvements.

\section{Conclusions}\label{sec:Conclusions}

In this paper, we propose a magnetohydrodynamic multi-fluid model to simulate the fluid dynamics of the escaping atmosphere of the hot Neptune GJ436 b. The primary goal is to explore the impact of the magnetic field on the decoupling of H$^+$ and H in the atmosphere. From the simulated results, although the decoupling between H$^+$ and H in the model is not very pronounced, the magnetic field does indeed have a certain influence on their decoupling. The occurrence of well-known wind zones and dead zones in the planet's atmosphere is noted under the influence of the magnetic field. Our work suggests that the decoupling between particles may be more pronounced in the dead zones and regions farther from the planet, and it strongly depends on the atmospheric mass loss rate. Subsequent cases of the decoupling of H$^+$ and O further illustrate the potential for the decoupling of heavy ions. Although our model is currently a simple isothermal model, it provides a preliminary framework for understanding potentially decoupled regions in the fluid dynamics of escaping atmospheres with global magnetic fields. This has implications for studying the decoupling of other components in similar environments.

\begin{acknowledgements}
\section{Acknowledgements} \label{sec:ackn}

We thank the anonymous reviewers for their constructive comments, which helped improve the manuscript. This work is Supported by the Strategic Priority Research Program of Chinese Academy of Sciences, grant No. XDB 41000000, the National Key R$\&$D Program of China (grant No. 2021YFA1600400/2021YFA1600402) and the National Natural Science Foundation of China (grants Nos. 11973082, 12288102, U2031111, 12233006, 42305136). The authors gratefully acknowledge the “PHOENIX Supercomputing Platform ” jointly operated by the Binary Population Synthesis Group and the Stellar Astrophysics Group at Yunnan Observatories, Chinese Academy of Sciences.
\end{acknowledgements}

\begin{appendix}
\section{Code verification}\label{sec:apdx}
\begin{figure}[htbp]
\plotone{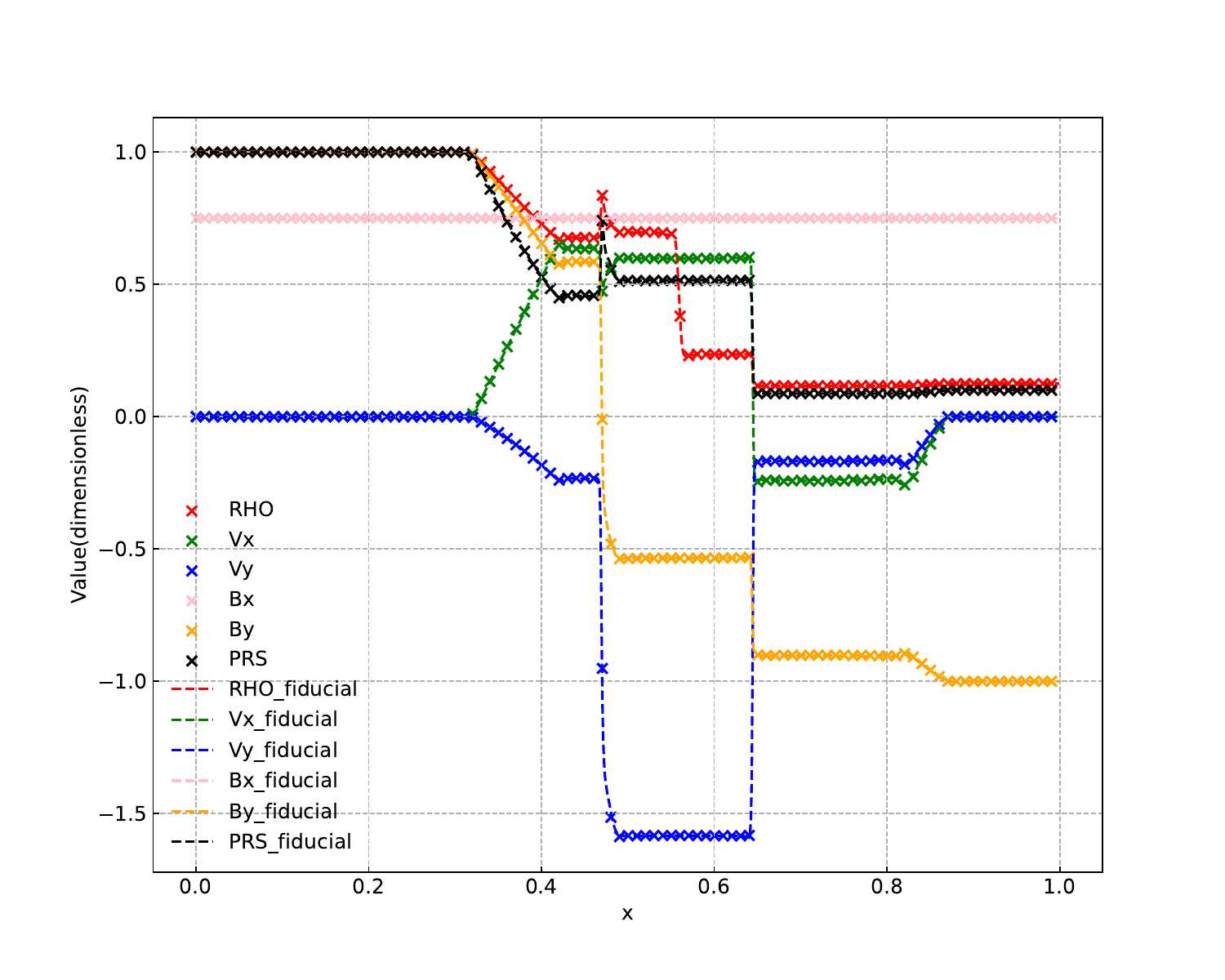}
\caption{
Here are the test results for the shock tube evolution at 0.1 time units, as presented in Table \ref{tab:SOD}. The dashed lines represent the simulation results using PLUTO as the benchmark for comparison. 'x' denotes the results from our program. The colors indicate different quantities: red for density, green for x-direction velocity, blue for y-direction velocity, pink for x-direction magnetic field, orange for y-direction magnetic field, and black for pressure. The z-direction magnetic field and velocity are not shown as they are both 0.
}

\label{fig:verification}
\end{figure}

\begin{table}[!t]
\begin{center}
\begin{threeparttable}[t]
\centering
\caption{Magnetohydrodynamic (MHD) Shock Tube Testing for the Magnetic Fluid Module}

\begin{tabular}{lllllll}
\toprule
  Side  &$\rho$ &$B_x$ &$B_y$ &$B_z$ &$Pressure$\\

\midrule
  Left  &1.0   &0.75 &1.0 &0.0 &1.0 \\
  Right &0.125 &0.75 &-1.0&0.0 &0.1 \\

\bottomrule

\end{tabular}\label{tab:SOD}

Here are the initial conditions for the shock tube. The spatial domain of the shock tube is one-dimensional, but vectors have three components. For $0 < x < 0.5$, the parameters correspond to the Left row, while for $0.5 < x < 1.0$, they correspond to the Right row. The initial conditions for the velocity vector are set to \textbf{0}. Additionally, the polytropic index $\gamma = 2.0$.

\end{threeparttable}
\end{center}
\end{table}

As our program is a magnetohydrodynamics (MHD) code developed based on PLUTO, it consists of fluid calculation module and MHD calculation module. The fluid module has been validated in \cite{Xing2023}. Here, we provide additional validation for the MHD part, specifically testing a dimensionless one-dimensional shock tube example with outflow boundary conditions. The initial conditions within the tube are divided into left and right halves, as specified in Table \ref{tab:SOD}. In this case, Figure \ref{fig:verification} illustrates a comparison between our results and those obtained using PLUTO. The agreement between our results (labeled as 'x') and PLUTO's results (dash lines) confirms the correctness of our MHD module.

\end{appendix}



\end{CJK*}
\end{document}